\documentclass[aps,preprint,epsf]{revtex4}

\usepackage{graphicx}% Include figure files

\begin{document}
\draft

\title{Thermodynamics, Structure, and Dynamics of Water \\
Confined between Hydrophobic Plates} 
\author{Pradeep Kumar,$^1$ Sergey V. Buldyrev,$^{1,2}$ Francis
  W.~Starr,$^{3}$ Nicolas~Giovambattista,$^{1,4}$ and H. Eugene
  Stanley$^1$} 

\address{$^1$Center for Polymer Studies and Department of Physics\\
 Boston University, 590 Commonwealth Avenue, Boston, MA 02215 USA\\
$^2$Department of Physics, Yeshiva University \\ 
500 West 185th Street, New York, NY 10033 USA \\
$^3$Department of Physics, Wesleyan University, Middletown, CT 06459
 USA\\ 
$^4$Department of Chemical Engineering, Princeton University, Princeton,
 NJ 08544 USA} 

\date{Submitted to Phys.~Rev.~E: 27 May 2005}

\begin{abstract}

We perform molecular dynamics simulations of 512 water-like molecules
that interact via the TIP5P potential and are confined between two
smooth hydrophobic plates that are separated by 1.10~nm. We find that
the anomalous thermodynamic properties of water are shifted to lower
temperatures relative to the bulk by $\approx 40$~K. The dynamics and
structure of the confined water resemble bulk water at higher
temperatures, consistent with the shift of thermodynamic anomalies to
lower temperature. Due to this $T$ shift, our confined water simulations
(down to $T = 220$~K) do not reach sufficiently low temperature to
observe a liquid-liquid phase transition found for bulk water at
$T\approx 215$~K using the TIP5P potential. We find that the different
crystalline structures that can form for two different separations of
the plates, $0.7$~nm and $1.10$~nm, have no counterparts in the bulk
system, and discuss the relevance to experiments on confined water.

\end{abstract}

\maketitle

\section{Introduction}

Despite the numerous accomplishments in water research to date
\cite{pablo-rev,Debenedetti03a,Debenedetti03b,braskin}, the topic
continues to be the subject of intense interest. In particular, water
confined in nanoscale geometries has garnered much recent attention due
to its biological and technological importance
\cite{zangi-rev,gelb}. Confinement can lead to changes in both
structural and dynamical properties caused by the interaction with a
surface and/or a truncation of the bulk correlation length. Moreover
these changes depend on whether the interactions of water with the wall
particles are hydrophilic or hydrophobic~\cite{chandler1}.  One of the
motivations for studying two different kinds of interactions arises from
studies of protein folding, since the folding of a protein is influenced
by its hydrophobic and hydrophilic interactions with water
\cite{protein}.

It is not clear exactly how the dynamics of liquids depends on the
nature of the confining surfaces. The behavior may change depending on
the surface morphology. Simulations of simple liquids show that the
dynamics typically {\it slow down\/} near a non-attractive rough surface
while the dynamics {\it speed up\/} near a non-attractive smooth surface
\cite{kb1}.  A slowing down of water dynamics near a hydrophilic
surface has been experimentally observed \cite{bellisent}.  Water
confined in Vycor \cite{Chen95,Chen94} has at least two different
dynamical regimes arising from the slow dynamics of water near the
surface and fast dynamics of water far away from the surfaces
\cite{gallo1,Spohr99,Gallo00,Hartnig00,Gallo99,Gallo00b}.

One anomaly hypothesized to occur in supercooled water is the emergence
of a phase transition line separating liquid states of different
densities. This phenomenon is called a liquid-liquid (LL) phase transition
\cite{pses92,ms98,poole1,speh,slhp,Sciortino03}. A LL transition has
been seen in a variety of simulation models of water
\cite{masako,pses92,hpss}, but is difficult to observe experimentally
due to the propensity of ice to nucleate at temperatures where a
transition is expected. Nonetheless, indirect evidence of a transition
has been found \cite{Mishima98,ms98,chenpaper,ChenPRVT,richert}. Studies
of some simple models of liquid also show a LL phase
transition \cite{Ja98,Bul02,Fr01,fran02,Ku04}.

Bulk water simulations using the TIP5P potential
\cite{jorgensen1,jorgensen2} indicate the presence of a LL phase
transition ending in a second critical point at $T\approx 217$~K and
$\rho\approx 1.17$~g/cm$^3$ \cite{masako,paschek1}. A LL phase
transition has been suggested based on simulations using the ST2
potential confined between smooth plates \cite{meyer}. A
liquid-to-amorphous transition is seen in simulations using the TIP4P
potential \cite{tip4p,tip4p-tanaka1,tip4p-tanaka2} confined in carbon
nanotubes \cite{koga2}.  Recent theoretical work \cite{truskett}
suggests that hydrophobic confinement suppresses the LL transition to
lower $T$. Here we aim to determine how confinement between smooth
hydrophobic walls affects the location of the the LL critical point as
well as the overall thermodynamic, dynamic and structural properties.

The freezing of water in confined spaces is also interesting. On one
hand, recent experimental studies of water confined in carbon nanopores
show that water does not crystallize even when the temperature is cooled
down to 77~K \cite{bellisent}. On the other hand, computer simulation
studies show that models of water can crystallize into different
crystalline forms when confined between surfaces
\cite{koga1,zangi1,zangi2,tanaka2,koga3}. For example, monolayer ice was found
in simulations using the TIP5P model of water \cite{zangi1}. By applying
an electric field along lateral directions (directions perpendicular to
the confinement direction) another crystalline structure for three
molecular layers of water confined between two silica plates was found
\cite{zangi2}.  Also, bilayer hexagonal ice was found in simulations
using the TIP4P model \cite{koga1}. In general these simulations predict
a variety of polymorphs in confined spaces, but the crystalline
structures found have yet to be observed in experiments.

This paper is organized as follows: In Sec.~II, we provide details of
our simulations and analysis methods. Simulation results for the liquid
state are provided in Secs.~III, IV, and V. The crystal states are
discussed in Sec.~VI, and we conclude with a brief summary in Sec.~VII.

\section{Simulation and Analysis Methods}

We perform molecular dynamics (MD) simulations of a system composed of
$N=512$ water-like molecules confined between two smooth walls.  The
molecules interact via the TIP5P pair potential \cite{jorgensen1} which,
like the ST2 \cite{st2} potential, treats each water molecule as a
tetrahedral, rigid, and non-polarizable unit consisting of five point
sites. Two positive point charges of charge $q_H=0.241e$ (where $e$ is
the fundamental unit of charge) are located on each hydrogen atom at a
distance $0.09572$~nm from the oxygen atom; together they form an $HOH$
angle of $104.52^{\circ}$. Two negative point charges ($q_e=-q_H$)
representing the lone pair of electrons ($e^-$) are located at a
distance $0.07$~nm from the oxygen atom. These negative point charges
are located in a plane perpendicular to the $HOH$ plane and form an
$e^-Oe^-$ angle of $\cos^{-1}(1/3)=109.47^{\circ}$, the tetrahedral
angle.  To prevent overlap of molecules, a fifth interaction site is
located on the oxygen atom, and is represented by a Lennard-Jones (LJ)
potential with parameters $\sigma_{OO}=0.312$~nm and
$\epsilon_{OO}=0.6694$~kJ/mol.

The TIP5P potential accurately reproduces many water anomalies when no
confinement is present \cite{masako}. For example, it accurately
reproduces the density anomaly at $T=277$~K and $P=1$~atm.  Its
structural properties compare well with experiments
\cite{masako,jorgensen1,jorgensen2,sorenson}. TIP4P and TIP5P are known
to crystallize \cite{masako,Ohmine} within accessible computer
simulation time scales; TIP5P shows a ``nose-shaped'' curve of
temperature versus crystallization time~\cite{masako}, a feature found
in experimental data on water solutions \cite{baez}. TIP5P simulations
also show a van der Waals loop in the $P-\rho$ plane at the lowest $T$
accessible with current computation facilities \cite{masako}. This loop
indicates the presence of a first-order LL transition.
Ref.~\cite{masako} estimates that a LL-transition line ends in a LL
critical point $C'$ located at $T_{C'}=217 \pm3$~K, $P_{C'}=340 \pm
20$~MPa, and $\rho_{C'}=1.13 \pm 0.04$~g/cm$^3$.

\begin{figure}[htb]
\includegraphics[width=12cm]{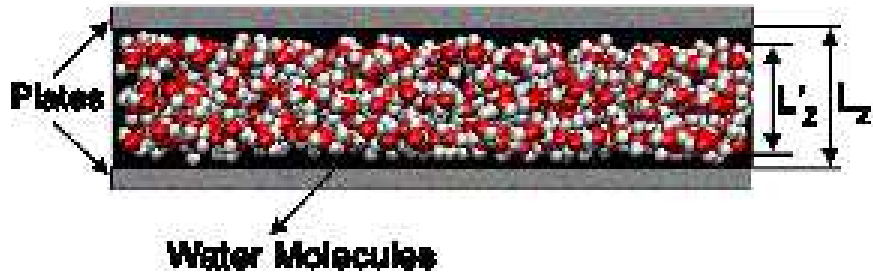}
\caption{Schematic of the simulated system. Water molecules are confined
   between two smooth hydrophobic plates in an $xy$ plane, separated by
   $L_z = 1.1$~nm.  The figure also indicates $L_z'$, the effective
   length of confinment of the water molecules.}
\label{scheme}
\end{figure}

In our simulation, water molecules are confined between two infinite
smooth planar walls, as shown schematically in Fig.~\ref{scheme}.  The
walls are located at $z=\pm0.55$~nm, corresponding to a wall-wall
separation of $1.1$~nm, which results in $\approx 2-3$ layers of water
molecules.  Periodic boundary conditions are used in the $x$ and $y$
directions, parallel to the walls.

The interactions between water molecules and the smooth walls are
designed to mimic solid paraffin \cite{lee2} and are given by
\cite{hansen}
\begin{equation}
U\left(\Delta z\right) =
4\epsilon_{\rm OW}\left[\left(\frac{\sigma_{\rm OW}}{\Delta
z}\right)^9-\left(\frac{\sigma_{\rm OW}}{\Delta z}\right)^3\right].
\label{Udz}
\end{equation}
Here $\Delta z$ is the distance from the oxygen atom of a water molecule
to the wall, while $\epsilon_{\rm OW} = 1.25$ kJ/mole and $\sigma_{\rm OW}=
0.25$ nm are potential parameters. The same parameter values were used
in previous confined water simulations \cite{lee1,lee2}.  

We perform simulations for $56$ state points, corresponding to seven
temperatures $T=220$, 230, 240, 250, 260, 280, and 300~K, and eight
densities $\rho=0.80$, 0.88, 0.95, 1.02, 1.10, 1.17, 1.25, and
1.32~g/cm$^3$ \cite{footnoteRealLz}. The range of density values takes
into account the fact that the water-wall interactions prevent water
molecules from accessing a space near the walls. Our determination of
$\rho$ is discussed in detail in the next section. The raw ``geometric''
densities used are $\rho=0.60$, 0.655, 0.709, 0.764, 0.818, 0.873, 0.927
and 0.981~g/cm$^3$.

For each state point, we perform two independent simulations to improve
the statistics.  We control the temperature using the Berendsen
thermostat with a time constant of $5$ ps \cite{berend} and use a
simulation time step of $1$~fs, just as in the bulk system
\cite{masako}. For long-range interactions we use a cutoff of $0.9$ nm
\cite{jorgensen1}.

We calculate the lateral pressure $P_{\|}=(P_{xx}+P_{yy})/2$ using the
virial expression for the $x$ and $y$-directions \cite{virial}. We
obtain the pressure along the transverse direction, $P_{\perp}$ by
calculating the total force $F_{\rm wall}$ perpendicular to the wall
\cite{meyer},
\begin{equation}
P_{\perp} = \frac{F_{\rm wall}}{L_xL_y} = \frac{|\sum_{i=1}^N{\bf
    F}_{i,{\rm wall}}|}{L_xL_y}.
\end{equation}
Here, ${\bf F}_{i,{\rm wall}}$ is the force produced by oxygen atom of water molecule
$i$ on the wall. Hydrogen atoms do not interact with the wall. In agreement with the simulations of Ref.~\cite{lee1}
using the TIP4P model for water and the water-wall interaction given by
Eq.~(\ref{Udz}), we find that the hydrogen atoms of the water molecules
near a wall tend to face away from the wall, forming bonds with other
molecules. 

\section{Properties of TIP5P Confined Water}

\subsection{Transverse Density Profile}
\label{sec:rhoz} 

\begin{figure}[htb]
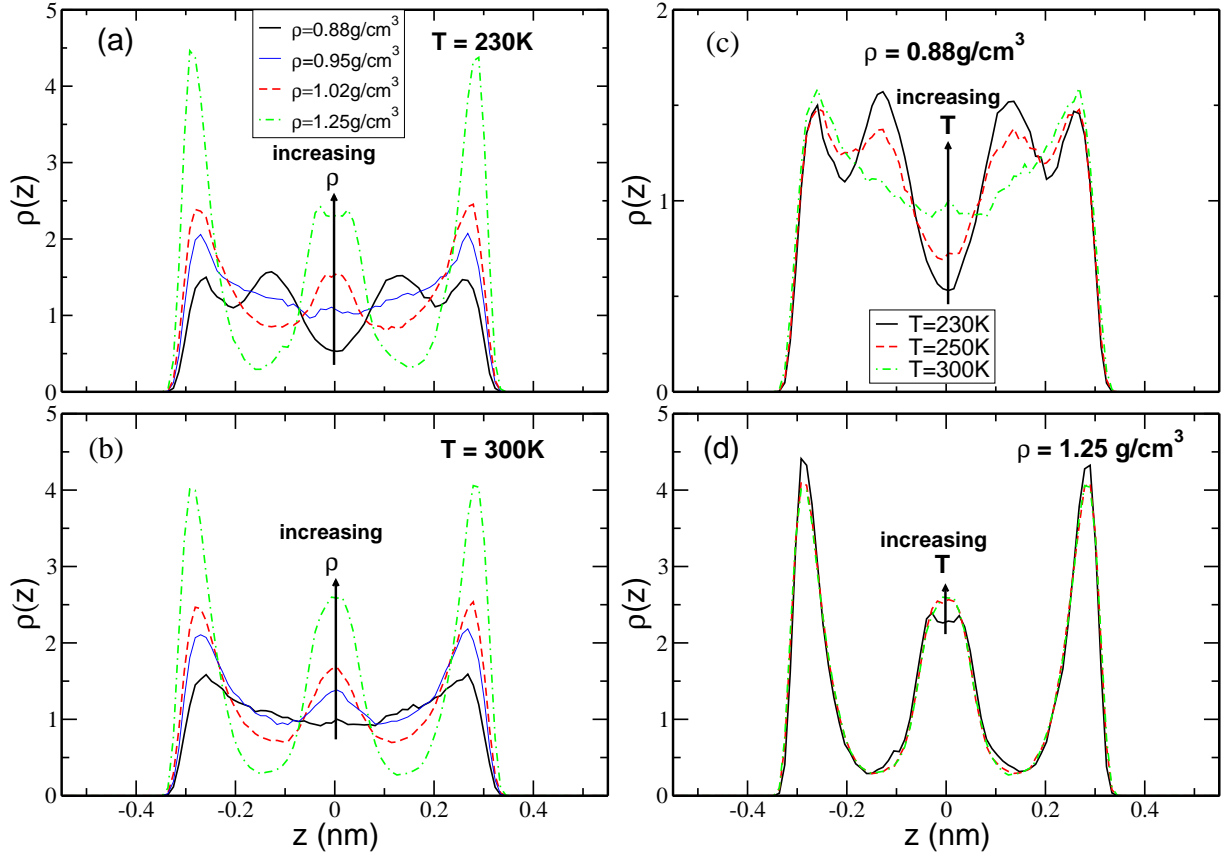

\includegraphics[width=8cm]{fig2a.eps} 
\includegraphics[width=8cm]{fig2c.eps} 
\includegraphics[width=8cm]{fig2b.eps} 
\includegraphics[width=8cm]{fig2d.eps} 
\caption{Density of water $\rho(z)$ along the confinement direction for
   different densities at constant temperatures (a) $T = 230$~K and (b)
   $T= 300$~K.  (c) Density of water along the confinement direction for
   different temperatures at constant densities (c) $\rho = 0.88$~
   g/cm$^3$ and (d) $\rho = 1.25$~g/cm$^3$. See discussion in text.}
\label{rho-z}
\end{figure}

One of the problems when dealing with liquids in a confined geometry is
how to define the density in a consistent way.  Using a geometric
definition $\rho\equiv Nm/L_xL_yL_z$ (where $m$ is the water molecule
mass) underestimates the effective density since the repulsive
interactions with the walls prevent molecules from coming too close to
the walls.  Hence we want to quantify the effective distance $L'_z$
perpendicular to the walls accessible to the water molecules (Fig.~1),
and thus obtain a definition for $\rho$ which is more readily comparable with
the density of a bulk system. To estimate $L'_z$, we calculate the
density profile $\rho(z)$ defined as the density of oxygen centers at $z$, shown in Fig.~\ref{rho-z} for different
temperatures and densities. In all cases studied, we observe that
molecules cannot access the total available space between the walls, and
that the accessible space along the transverse direction does not
strongly depend on $T$ and $\rho$. Hence we estimate
\begin{equation}
L'_z=L_z-\frac{\sigma_{\rm OW}+\sigma_{\rm OO}}{2}=0.819~{\rm nm}
\end{equation}
independent of $T$ and $\rho$; this leads to the effective density
\begin{equation}
\rho\equiv {Nm\over L_xL_yL'_z}.
\end{equation}

Figure~\ref{rho-z}(a) shows the effect on $\rho(z)$ of changing $\rho$
at $T=230$~K. Since the typical oxygen-oxygen separation for nearest
neighbor in bulk water is $0.28$~nm, for the effective wall separation
$L'_z=0.819$~nm one would expect that at most three water layers can be
accommodated between the walls.  At $\rho=1.02$~g/cm$^3$ and
$\rho=1.25$~g/cm$^3$, $\rho(z)$ shows three clear
maxima indicating the presence of a trilayer liquid.  The two maxima
next to the walls are the result of water-wall interaction. As density
decreases below $\rho=1.02$~g/cm$^3$, the central maximum becomes nearly
uniform and, at density $\rho=0.95$~g/cm$^3$, only the two maxima
located next to the walls remain. This density corresponds to the least
structured liquid. Upon further expansion, the structure of the liquid
starts to increase since the bilayer splits into two sublayers for the
lowest density $\rho=0.88$g/cm$^3$. As we will see in the next section,
$T=230$~K is below the temperature of maximum density ($T_{\rm
MD}$). Hence the increase in the structure upon expansion corresponds to
the anomalous decrease in entropy upon expansion found in bulk water
below the $T_{\rm MD}$,
\begin{equation}
\left(\frac{\partial S}{\partial V}\right)_T = \left(\frac{\partial
P}{\partial T}\right)_V = -\left(\frac{\partial V}{\partial
  T}\right)_P\left(\frac{\partial P}{\partial
  V}\right)_T\leq 0.
\end{equation}

Figure~\ref{rho-z}(b) shows the effect on $\rho(z)$ of changing $\rho$
at higher $T$, $T=300$~K. The densities in Fig.~\ref{rho-z}(b) are the
same as those shown in Fig.~\ref{rho-z}(a). Similar to findings at
$T=230$~K, (i) the liquid at $T=300$~K and $\rho=1.25$~g/cm$^3$ is
characterized by a trilayer structure and (ii) reducing the density
transforms the trilayer liquid into a bilayer liquid at
$\rho=0.88$~g/cm$^3$.  However, comparison of Fig.~\ref{rho-z}(a) and
\~ref{rho-z}(b) shows that, at the lowest $\rho$ studied, the layers of
the bilayer structured liquid at $T=300$~K do not split into two
sub-layers, as is the case of $T=230$~K since $T=300$ lies above the
$T_{\rm MD}$. In fact, we find that at $T=230$~K the sublayers present
at $\rho=0.88$~g/cm$^3$ merge into a single layer at
$\rho=0.95$~g/cm$^3$, and the resulting $\rho(z)$ resembles that shown
in Fig.~\ref{rho-z}(b) at $\rho=0.88$~g/cm$^3$.

The effect of changing temperature at low density is shown in
Fig.~\ref{rho-z}(c). As discussed above, at low density and low $T$ we
observe a bilayer liquid where the two layers split into two
sublayers. Increasing $T$ smoothes features of the density profile,
namely (i) the splitting of the sublayers disappears, and (ii) the
minimum of $\rho(z)$ at $z=0$~nm becomes less pronounced since
increasing $T$ increases the entropy. While a
bilayer~$\rightarrow$~trilayer liquid crossover is found upon isothermal
compression, isochoric heating does not have such an effect.

Figure~\ref{rho-z}(d) shows the effect of changing $T$ at the high
density, $\rho=1.25$~g/cm$^3$. At high $\rho$, the molecules are
not able to displace perpendicular to the walls and the density profile
is almost $T$ independent. Indeed, the liquid has a trilayer structure
at all $T$ studied.
 
\subsection{Low Temperature Phase Diagram}

Next we test how confinement affects the location in the $P-T$ phase diagram of the LL phase transition line
and the second critical point found in bulk water simulations using the
TIP5P model \cite{masako,paschek1}. To determine the phase behavior, we
evaluate $P$ as a function of $\rho$. If there is a second LL phase
critical point, then it should manifest itself in a van der Waals
``loop'' along isothermal paths at low $T$.

\begin{figure}[htb]
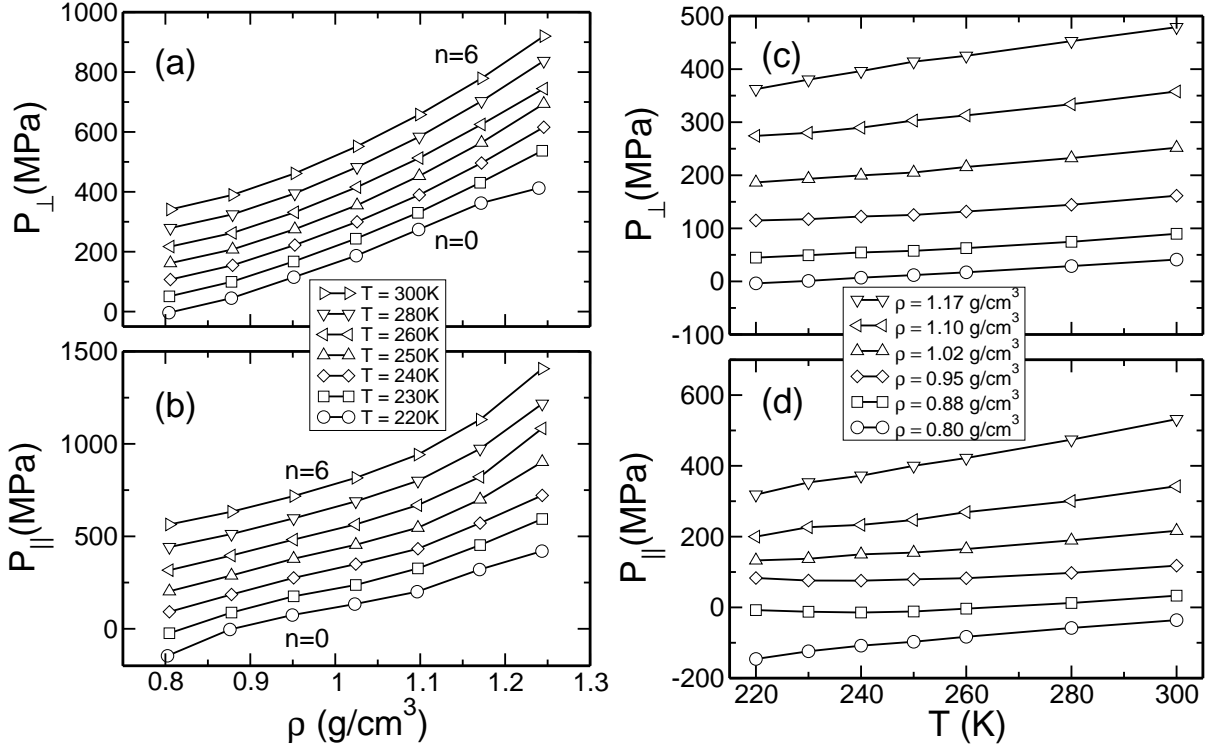

\includegraphics[width=8.cm]{fig3ab.eps} 
\includegraphics[width=7.8cm]{fig3cd.eps}

\caption{(a) Transverse pressure $P_{\perp}$ (perpendicular to the
   walls) and (b) lateral pressure $P_{\|}$ (parallel to the walls) as
   functions of density $\rho$ for all simulated $T$. For clarity, each
   curve is shifted by $n \times 50 $MPa for $P_{\perp}$ for and $n
   \times 100 $MPa for $P_{\|}$. While $P_{\perp}$ shows no infection for
   all $T$, $P_{\|}$ starts to flatten at $\rho \approx 0.87$~g/cm$^3$
   for the lowest $T$ presented. This is consistent with the possibility
   of a second critical point at lower than $T=220$~K. (c) $P_{\perp}$
   and (d) $P_{\|}$ as a function of temperature for different
   densities.  Only $P_\|$ displays minima, indicating the presence of a
   $T_{\rm MD}$.}
\label{PhaseD}
\end{figure}

Unlike bulk liquid systems, one must be careful to interpret separately
the results involving the lateral pressure $P_{\|}$ and those involving
the transverse pressure $P_{\perp}$, since the thermodynamic averages of
these quantities will usually be different. Phase separation will only
be apparent in $P_{\|}$, since the separation of the plates is too small
to allow for the existence of two distinct phases in the transverse
direction.  In Figs.~\ref{PhaseD}(a) and \ref{PhaseD}(b), we show
$P_{\perp}$ and the lateral pressure $P_{\|}$ as functions of density
along all seven isotherms.

\begin{figure}[htb]
\includegraphics[width=10cm]{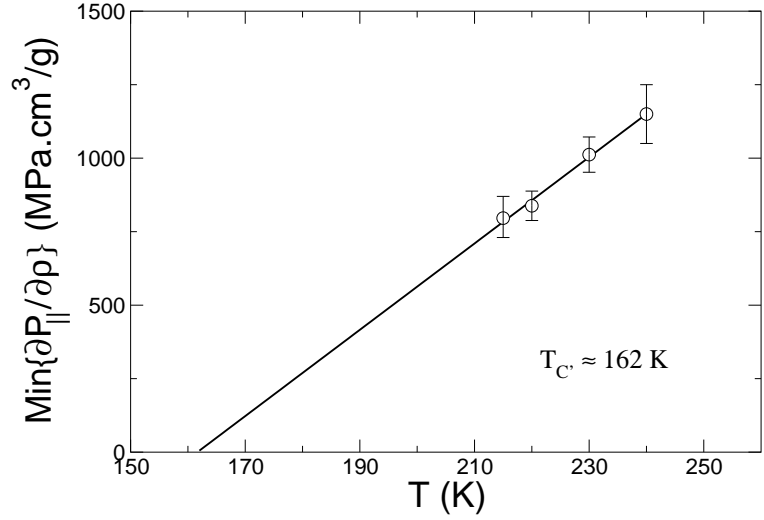}
\caption{Minimum values of ($\partial P_\|/\partial \rho)_T$ as a
function of temperature obtained from Fig.~\ref{PhaseD}(d). The critical
temperature $T_{C'}$ of the critical point of the LL transition is
crudely estimated by extrapolation.}

\label{slope-isochore}
\end{figure}

Since there can be no phase separation in the transverse direction,
$P_{\perp}$ is a monotonically increasing function of the density
(Fig.~\ref{PhaseD}(a)). $P_{\|}$ is also a monotonic increasing
function of $\rho$, but as $T$ decreases, the isotherms of $P_{\|}$
become ``flatter'' in the region near $\rho \approx 0.95$~g/cm$^3$
(Fig.~\ref{PhaseD}(b)).  The presence of this region is consistent with
the possible existence of a van der Waals loop at lower $T$ and,
therefore, is consistent with the possible existence of a LL phase
transition line ending at a second critical point at a value of $T$
lower than 220~K, the lowest simulated temperature. At 220~K no phase
separation occurs, consistent with the simulations of bulk water where
the same TIP5P potential gives a second critical point with $T_{C'}=217
\pm 3$~K \cite{masako}.  While we are unable to simulate the
temperatures below $T_{C'}$ of the bulk system, comparison of the lowest
$T$ isotherm with that of Ref.~\cite{masako,paschek1} suggests that we
will need to go well below the bulk $T_{C'}$ to see phase separation.
Thus our results suggest that the presence of hydrophobic walls shifts a
possible second critical point to lower $T$. Along an isothermal path,
the critical point can be located by the point where the slope
$(\partial P/\partial\rho)_T$ and curvature
$(\partial^2P/\partial\rho^2)_T$ simultaneously equal zero. We can
estimate this point by plotting the values of the minimum slopes along
each isotherm and extrapolating the slopes to find the $T$ at which the slope is zero.
This estimate yields critical temperature $T_{C'}\approx 162 \pm
20$~K~(Fig.~\ref{slope-isochore}).

\begin{figure}[htb]
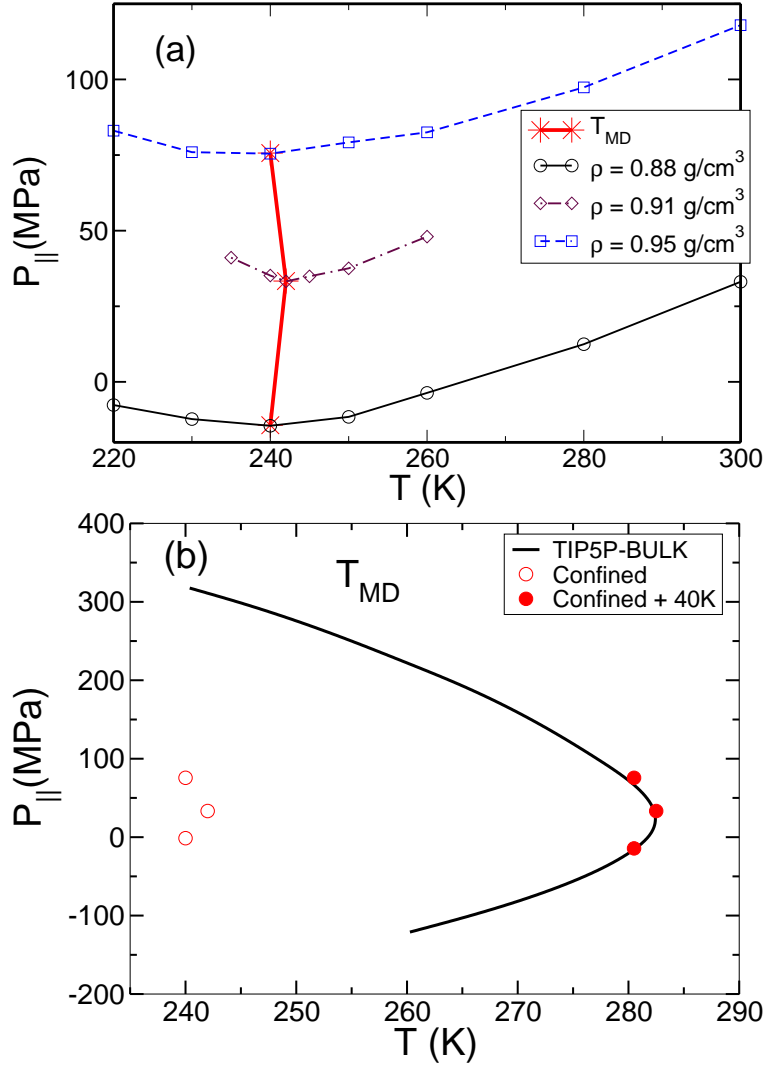

\includegraphics[width=10cm]{fig5a.eps}
\includegraphics[width=10cm]{fig5b.eps}
\caption{A comparison of $T_{\rm MD}$ line for bulk and confined
water. (a) Isochores corresponding to the densities $\rho=0.88$g/cm$^3$,
$\rho=0.91$g/cm$^3$ and $\rho=0.95$g/cm$^3$ which have a minimum. The
locus of the points where $(\partial P_\|/\partial T)_V=0$ denotes the $T_{\rm
MD}$ (shown as asterisk). (b) If the $T_{\rm MD}$ for confined
water (open circles) were shifted by $40$~K, it would overlap the $T_{\rm MD}$ for bulk
water (filled circles).}
\label{tmd-plot}
\end{figure}

Figures~\ref{PhaseD}(c) and \ref{PhaseD}(d) show the $T$-dependence of
$P_{\perp}$ and $P_{\|}$ for different densities. $P_{\perp}(T)$ is a
monotonic function of $T$ for all $\rho$. Similar behavior is observed
for $P_{\|}(T)$ at large $\rho$. However, for $0.88$~g/cm$^3$~$\leq \rho
\leq 1.17$~g/cm$^3$, the isochores in the $P_{\|}-T$ plane display
minima, indicating the presence of a $T_{\rm MD}$ line, defined as the
locus of points where $(\partial P/\partial T)_{V}=0$ \cite{hansen}. For
$T<T_{\rm MD}$, water confined between hydrophobic walls is anomalous,
i.e., it becomes less dense upon cooling. A $T_{\rm MD}$ line has also
been found in TIP5P bulk water simulations \cite{masako}.  Comparison of
Fig.~\ref{PhaseD}(c) and Fig.~2(a) of Ref.~\cite{masako} shows that the
$T_{\rm MD}$ locus in confined water shifts to lower $T$. We also plot
the $T_{\rm MD}$ for bulk water (from Ref.~\cite{masako}) and confined
water in Fig.~\ref{tmd-plot}. A $+40$~K temperature shift in the $T_{\rm
MD}$ of confined water overlaps these loci. Thus the effect of the
hydrophobic walls in our system seems to be to shift the $P-T$ phase
diagram by $\Delta T\approx-40\pm 5$~K with respect to bulk water. This
is consistent with the second critical point shifting to lower $T$.

\begin{figure}[htb]
\includegraphics[width=12cm]{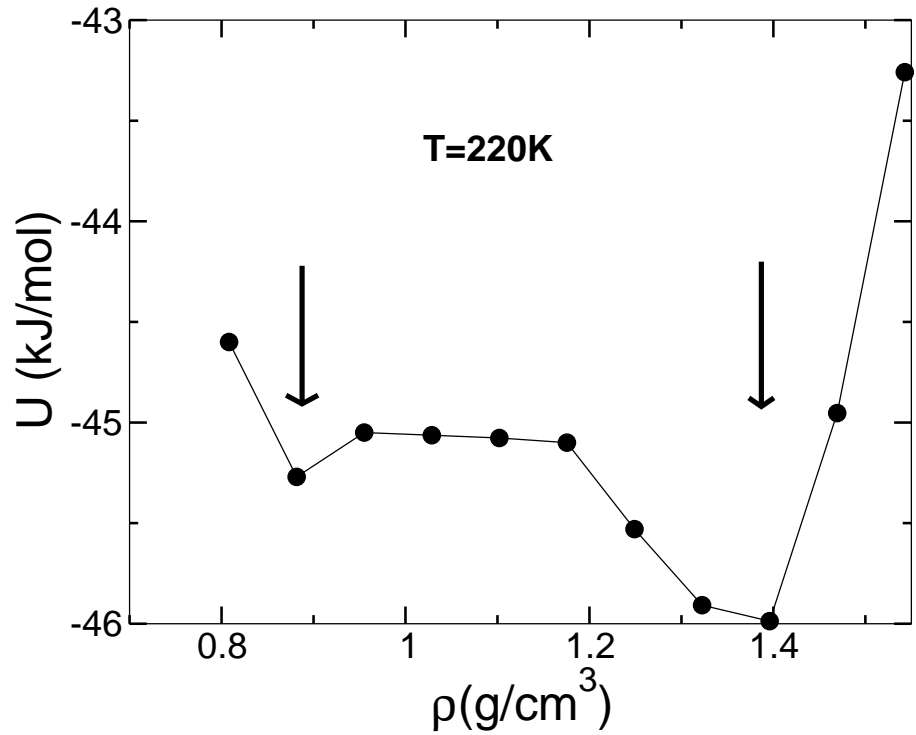}
\caption{Potential energy as a function of $\rho$ for the lowest
   temperature we simulate, indicating the possibility of two stable
   liquid states at $\rho=0.88$g/cm$^3$ and $\rho=1.39$g/cm$^3$. The two
   minima are indicated by vertical arrows.}
\label{u-rho}
\end{figure}

Figure~\ref{u-rho} shows the calculated potential energy for the lowest
simulated temperature $T=220$~K. We note two minima for the densities
around $\rho=0.88$g/cm$^3$ and $\rho=1.39$g/cm$^3$ respectively. Since
the free energy $F$ is given by
\begin{equation}
 F = K + U - TS,
\end{equation}
where $K$, $U$, and $S$ are the kinetic energy, potential energy and
entropy respectively, at small $T$ an extremum in $U$ suggests an
extremum in $F$. Hence the emergence of two minima at small $T$ further
supports the possibility of two stable liquids at low and high densities
respectively.

\section{Static Structure}

\subsection{Radial Distribution Function}

In Sec.~II we studied the structure of water along the direction
perpendicular to the walls. To aid in comparing the structural
properties with those of bulk water, we next focus on the lateral
oxygen-oxygen radial distribution function (RDF) defined by
\begin{equation}
g_{\|}(r) \equiv \frac{1}{\rho^2 V}\sum_{i \ne j} \delta(r-r_{ij})
\left[\theta \left(|z_i-z_j|+\frac{\delta z}{2}\right) -\theta
\left(|z_i-z_j|-\frac{\delta z}{2}\right)\right].
\end{equation}
Here $V$ is the volume, $r_{ij}$ is the distance parallel to the walls
between molecules $i$ and $j$, $z_i$ is the $z$-coordinate of the oxygen
atom of molecule $i$, and $\delta(x)$ is the Dirac delta
function. The Heaviside functions, $\theta(x)$, restrict the sum to a
pair of oxygen atoms of molecules located in the same slab of thickness
$\delta z = 0.1$~nm.  The physical interpretation of $g_{\|}(r)$ is that
$g_{\|}(r) 2 \pi r dr \delta z$ is proportional to the probability of
finding an oxygen atom in a slab of thickness $\delta z$ at a distance
$r$ parallel to the walls from a randomly chosen oxygen atom. In a bulk
liquid, this would be identical to $g(r)$, the standard RDF.

\begin{figure}[htb]
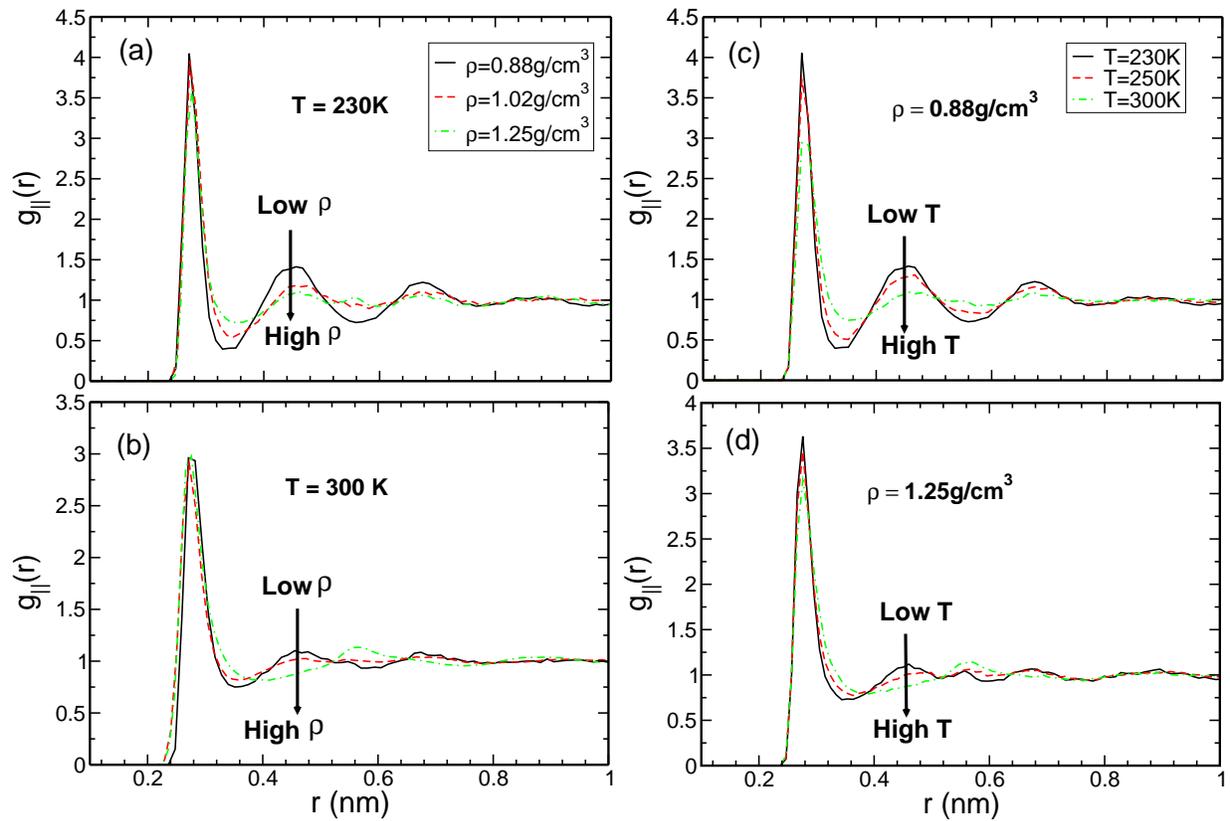

\includegraphics[width=8cm]{fig7a.eps}
\includegraphics[width=8cm]{fig7c.eps}
\includegraphics[width=8cm]{fig7b.eps}
\includegraphics[width=8cm]{fig7d.eps}
\caption{Lateral oxygen-oxygen radial distribution function $g_{||}(r)$
   for different densities at a constant temperature (a) $T$ = $230K$
   and (b) $T = 300$~K; $g_{||}(r)$ for different temperatures at a
   constant density (c) $\rho$ = $0.88$~g/cm$^3$ and (d) $\rho = 1.25$~g/cm$^3$.}
\label{gxy}
\end{figure}

Figure~\ref{gxy}(a) shows the effect on $g_{\|}(r)$ of increasing $\rho$
at low $T$. At low $\rho$, the bilayer liquid is characterized by a RDF
that resembles the RDF of bulk water at $\rho=0.99$~g/cm$^3$
\cite{finneyPRL}, with maxima at $r \approx 0.28$, 0.45, and 0.67~nm.
The well-defined maxima and minima in both $\rho(z)$
(Fig.~\ref{rho-z}(a)) and $g_{\|}(r)$ (Fig.~\ref{gxy}(a)) indicate that
at low-$T$ and low-$\rho$ the liquid is highly structured, with a
structure parallel to the walls similar to corresponding bulk liquid
water at low density \cite{soperRicci}.  As $\rho$ increases, the liquid
becomes less structured as indicated by the decreasing height of the
second and third peaks of $g_{\|}(r)$, and by the disappearance of the
sublayers in $\rho(z)$ (Fig.~\ref{rho-z}(a)). A comparison of
$g_{\|}(r)$ at $\rho=1.25$~g/cm$^3$ with the $g(r)$ for bulk water at
high density from Ref.~\cite{soperRicci} shows that the distributions
are very different. However, we find that if we calculate $g_{\|}(r)$
for only $-0.17$~nm $ \leq z \leq 0.17$~nm (which corresponds to the
location of the central layer), then $g_{\|}(r)$ at low-$\rho$ and
high-$\rho$ resemble $g(r)$ for bulk water at both low and high
densities respectively. Thus the evolution of the structure parallel to
the walls of the central layer mimics the structural changes when going
from low-density bulk water to high-density bulk water.

The effect of increasing the density at $T=300$~K is shown in
Fig.~\ref{gxy}(b).  The main effect of compression is to
``redistribute'' the molecules parallel to the walls.  At
$\rho=0.88$~g/cm$^3$, $g_{\|}(r)$ is similar to the distribution shown
in Fig.~\ref{gxy}(a) at the same density, which resembles $g(r)$ for
low-density bulk water. However, the maxima and minima in $g_{\|}(r)$
are less pronounced at $T=300$~K.  At intermediate $\rho$, $g_\|(r)$
becomes very close to $1$ for $r>0.45$~nm. As density increases up to
$\rho=1.25$~g/cm$^3$, a weak peak appears at $r\approx 0.9$~nm and a
maximum occurs at $0.55$~nm. Furthermore, the first maximum of
$g_{\|}(r)$ becomes wider and the first minimum shift towards $r\approx
0.4$~nm. The resulting $g_{\|}(r)$ at high $\rho$ and high $T$ has many
features of $g(r)$ for bulk water obtained experimentally at $T=268$~K
\cite{soperRicci}. For example, the oxygen-oxygen $g(r)$ in
Ref.~\cite{soperRicci} shows a weak peak at $r\approx 0.9$~nm, a clear
maximum at $0.61$~nm, and the first minimum is located at $r \approx
0.41$~nm.  Furthermore, a shoulder in the first maximum of $g(r)$
develops at high density which is consistent with the increase of the
width of the first peak of $g_{\|}(r)$ in Fig.~\ref{gxy}(b) as $\rho$
increases.

To complete the comparison of the structure of confined water with the
structure of bulk water, we also evaluate the $T$-dependence of
$g_{\|}(r)$.  In Fig.~\ref{gxy}(c), we show $g_{\|}(r)$ for various $T$
at $\rho=0.88$~g/cm$^3$.  The effects of increasing $T$ at low $\rho$
are similar to those observed in Fig.~\ref{gxy}(a) when increasing
$\rho$ at low $T$.  More specifically, the minima and maxima of
$g_{\|}(r)$ become less pronounced as $T$ increases and $g_{\|}(r)$
becomes flatter for $r > 0.5$~nm. We note that similar changes are also
found in $\rho(z)$ when (i) increasing $T$ at low $\rho$
(Fig.~\ref{rho-z}(a)) and (ii) increasing $\rho$ at low $T$
(Fig.~\ref{rho-z}(c)).

Figure~\ref{gxy}(d) shows the effect of $T$ on $g_{\|}(r)$ when the
density is fixed at $\rho=1.25$~g/cm$^3$. The effects of increasing $T$
at high $\rho$ (Fig.~\ref{gxy}(d)) are similar to the effects of
increasing $\rho$ at a high $T$ (Fig.~\ref{gxy}(b)).  More specifically,
the first minimum of $g_{\|}(r)$ shifts to a larger $r$ and becomes less
pronounced, while the second and third peaks located at $0.45$~nm and
$0.7$~nm, respectively, merge and form an intermediate peak at
$r=0.55$~nm.  This suggests that as $\rho$ increases at high $T$, the
preferred distance between second neighbors increases (parallel to the
walls) and local tetrahedral order decreases.  The emergence of a peak
at $0.55$~nm on heating at 1.25~g/cm$^{3}$ is in contrast to the
behavior of bulk water, where the disappearance of the peaks at 0.45 and
0.7 nm gives rise to nearly featureless behavior of $g(r)$ beyond the
first peak (see, e.g., Ref.~\cite{sslsss}).  Hence at high $T$,
confinement gives rise to structure that would not be present in bulk
systems, presumably because molecules orient relative to the walls.  It
is interesting to note that, as shown in Sec.~\ref{rho-z}, there is
almost no change in $\rho(z)$ with $T$, indicating that the
rearrangement of the molecules parallel to the walls has no effect, on
average, on the organization of molecules perpendicular to the walls.

\subsection{Static Structure Factor}

Next we calculate the lateral static structure factor $S_{\|}(q)$,
defined as the Fourier transform of the lateral radial distribution
function $g_{\|}(r)$,
\begin{equation}
S_{\|}\left(q\right) = \frac{1}{N}\sum_{i,j}\left<e^{i\vec{\bf
q}.(\vec{\bf r_i} - {\bf \vec{r_j}})}\right>,
\end{equation}
where the $q$-vector is the inverse space vector in the $xy$ plane and
$r$ is the projection of the position vector on the $xy$ plane. The
structure factor will be particularly useful for comparison with the
crystal structure in Sec.~\ref{sec:xtal}, where distinct Bragg peaks in
$S_{\|}\left(q\right)$ appear.

\begin{figure}[htb]
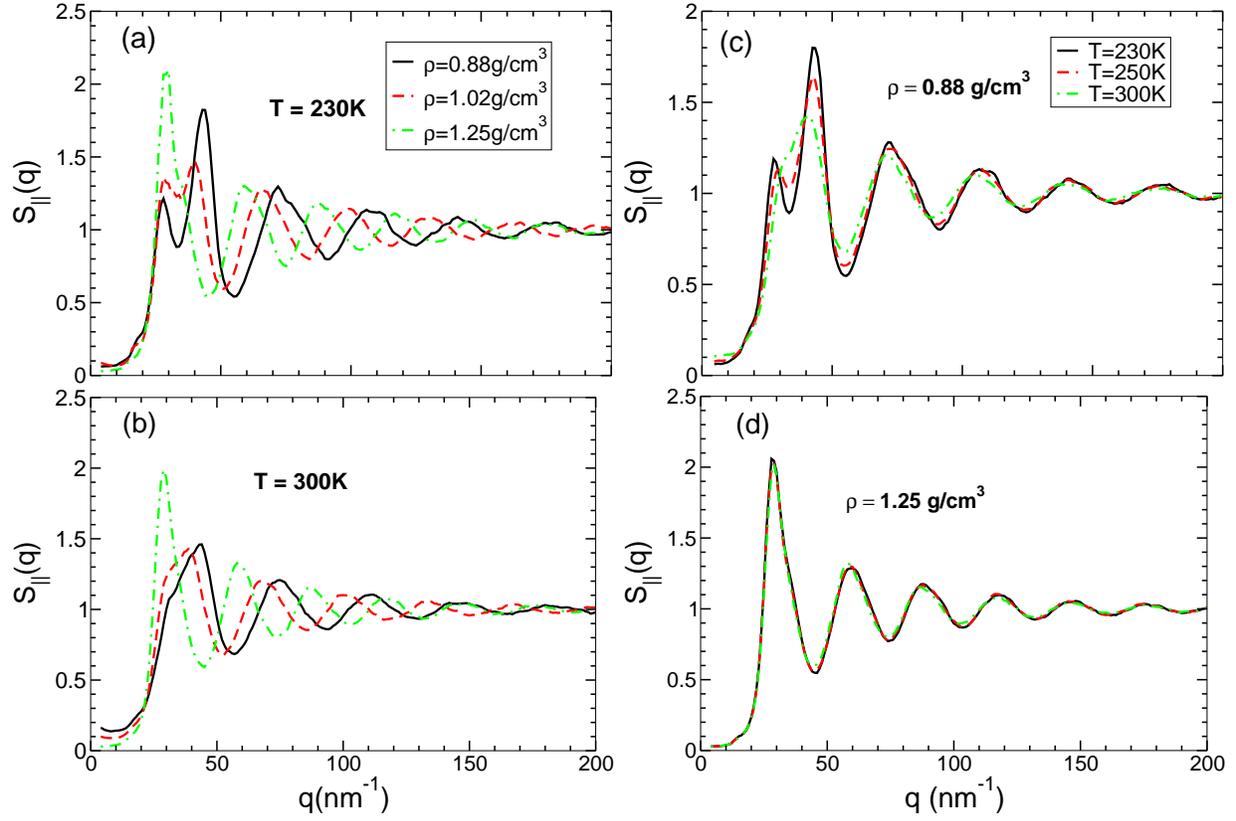

\includegraphics[width=8cm]{fig8a.eps}
\includegraphics[width=8cm]{fig8c.eps}
\includegraphics[width=8cm]{fig8b.eps}
\includegraphics[width=8cm]{fig8d.eps}
\caption{Lateral structure factor $S_{\|}(q)$ for different densities at
   a constant temperature (a) $T=230$~K and (b) $T = 300$~K; $S_{\|}(q)$
   for different temperatures a constant density (c)
   $\rho=0.88$~g/cm$^3$ and (d) $\rho=1.25$~g/cm$^3$.}
\label{sqxy}
\end{figure}

Figures~\ref{sqxy}(a) and \ref{sqxy}(b) show the effect of density on the
lateral structure factor $S_\|(q)$ for $T = 230$~K and $T = 300$~K.  The
structure factor of low-density and low-temperature confined water is similar
to bulk water.  The presence of a ``pre-peak'' in $S_{\|}\left(q\right)$ at
low $\rho$ can be attributed to the existence of pronounced tetrahedral order
in the low-temperature liquid.  At high $\rho$ or high $T$, this feature is
reduced, as the molecular order becomes less tetrahedral and core repulsion
dominates. This behavior is similar to bulk water, but with a shift to lower
densities. We show the evolution of $S_\|(q)$ as a function of $T$ for two
different densities $\rho=0.88$~g/cm$^3$ and $\rho=1.25$~g/cm$^3$ in
Figs.~\ref{sqxy}(c) and \ref{sqxy}(d). The structure of low-temperature water
for $\rho=0.88$~g/cm$^3$ is similar to low-density water. When the
temperature is increased, the repulsive region of the potential begins to
dominate and tetrahedrality is reduced. The first two peaks in the structure
factor merge to form a single peak (Fig.~\ref{sqxy}(c)). However, at high
density $\rho=1.25$~g/cm$^3$, a change is temperature does not change the
structure factor significantly (Fig.~\ref{sqxy}(d)). Similar behavior is seen
in bulk water \cite{francis}.

\section{Dynamics}

Thus far, we have seen that if a LL transition exists
for confined water, it is shifted to lower $T$ than for bulk water, and
that the tetrahedral order that gives rise to this behavior is also
suppressed.  Hence it is natural to consider whether the dynamic
properties of confined water exhibit the same temperature shift found
for the thermodynamic properties relative to bulk water. For example,
how is the maximum in diffusivity under pressure shifted under
confinement? To compare with the bulk system, we calculate the lateral
mean square displacement (MSD).  We can evaluate the diffusion
coefficient $D$ from the asymptotic behavior of the lateral MSD using
the Einstein relation
\begin{equation}
\langle r_{\|}^2\rangle = 2dDt,
\end{equation}
where $\langle r_{\|}^2\rangle$ is the mean square displacement parallel
to the walls over a given time interval $t$, and $d$ is the system
dimension \cite{benedek}.  Since we calculate the diffusion only in the
lateral directions, $d=2$.

\begin{figure}[htb]
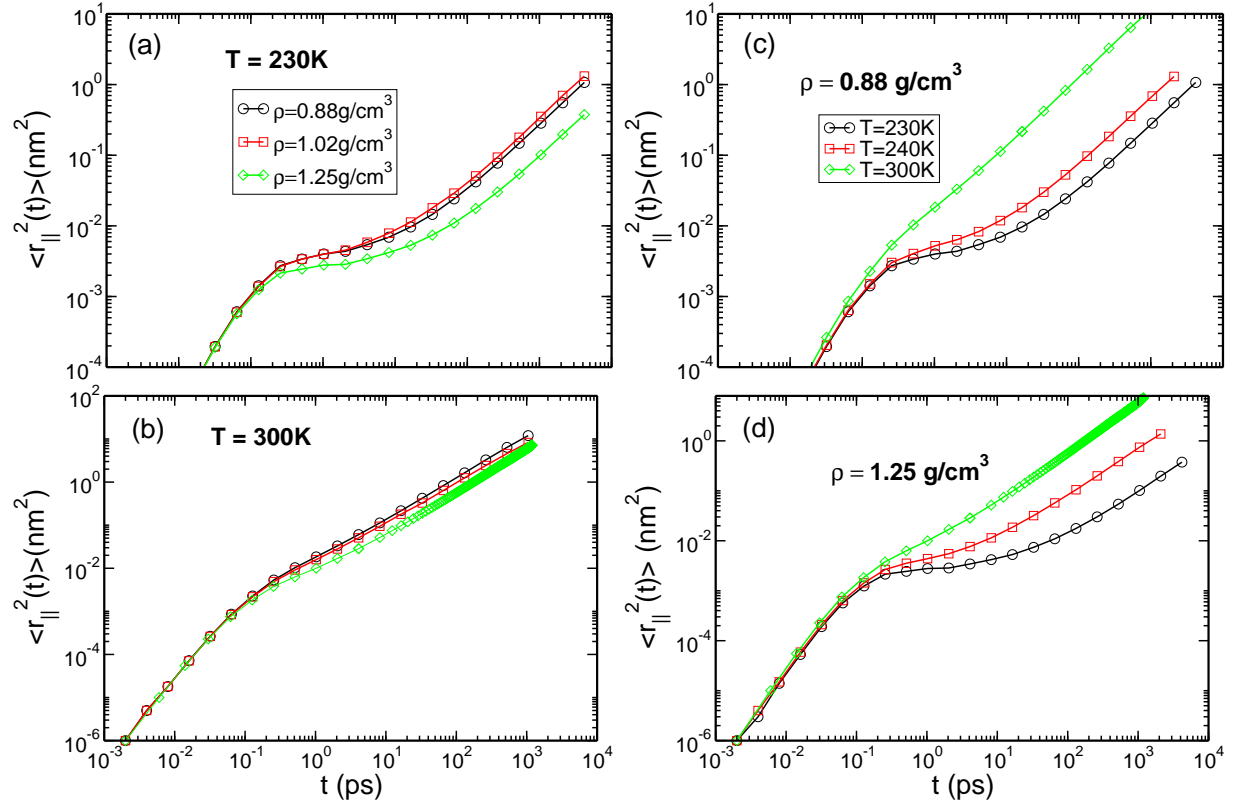

\includegraphics[width=8cm]{fig9a.eps}
\includegraphics[width=8cm]{fig9c.eps}
\includegraphics[width=8cm]{fig9b.eps}
\includegraphics[width=8cm]{fig9d.eps}
\caption{Lateral mean square displacement $\langle r_\|^2(t) \rangle$ for
   different densities at constant temperatures (a) $T=230$~K and (b)
   $T=300$~K.  $\langle r_\|^2(t) \rangle$ for different temperatures at
   constant densities (c) $\rho = 0.88$~g/cm$^3$ and (d) $\rho =
   1.25$~g/cm$^3$.}
\label{msd1}
\end{figure}

Figures~\ref{msd1}(a) and \ref{msd1}(b) show the dependence of the
lateral MSD on $\rho$ at fixed $T=230K$ and $T=300K$.  We also plot the
dependence of the lateral MSD on $T$ for two different temperatures in
Figs.~\ref{msd1}(c) and \ref{msd1}(d), using a log-log scale to
emphasize the different mechanisms seen on different time scales:
  
\begin{itemize}

\item[{(i)}] An initial ballistic motion, where the lateral MSD is a
      quadratic function of time, $\langle r_{\|}^2(t)\rangle\sim t^2$.

\item[{(ii)}] An intermediate ``flattening'' of the lateral MSD, due to
      the transient caging of molecules by their hydrogen bonded
      neighbors.  This effect is most noticeable at the lowest $T$
      studied, and does not occur at high $T$.

\item[{(iii)}] Long time scales on which particles diffuse randomly, and
      so $\langle r_{\|}^2(t)\rangle\sim t$.

\end{itemize}

\begin{figure}[htb]
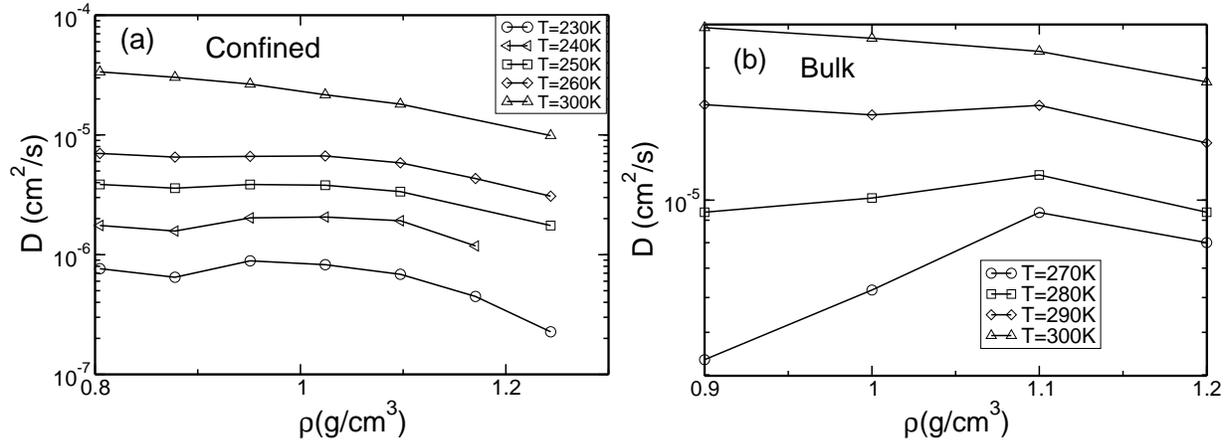

\includegraphics[width=8cm]{fig10a.eps}
\includegraphics[width=8cm]{fig10b.eps}
\caption{Diffusion constant $D$ as a function of density for various
   temperatures in (a) confined water and (b) bulk water. Confined water
   shows diffusion anomalies at lower temperatures than in bulk
   water. The first diffusion anomaly (the increase of diffusion with
   increase in density) in the confined system begins to appear when the
   temperature drops to $T\approx 250$~K, which is $\approx 40$~K lower
   than in bulk water (which displays diffusion anomalies at $T\approx
   290K$).}
\label{msd2}
\end{figure}

To determine whether there is an anomaly in the density dependence of
$D$, we plot $D$ along isotherms, in Fig.~\ref{msd2}(a) for confined
water and in Fig.~\ref{msd2}(b) for bulk water. For $T \lesssim 250$~K,
we find that $D$ has a maximum at $\rho\approx 1.05$~g/cm$^{3}$.  In
bulk water, a similar behavior is found, but at $T\approx 290$~K, $40$~K
higher than the confined system. Moreover, this shift in a dynamic
anomaly is consistent with the shift of thermodynamic anomalies.
Qualitatively, the maximum in $D$ can be understood as a competition
between weakening or breaking of hydrogen bonds under pressure
\cite{sns} (which increases $D$) and increased packing (which reduces
$D$).

\begin{figure}[htb]
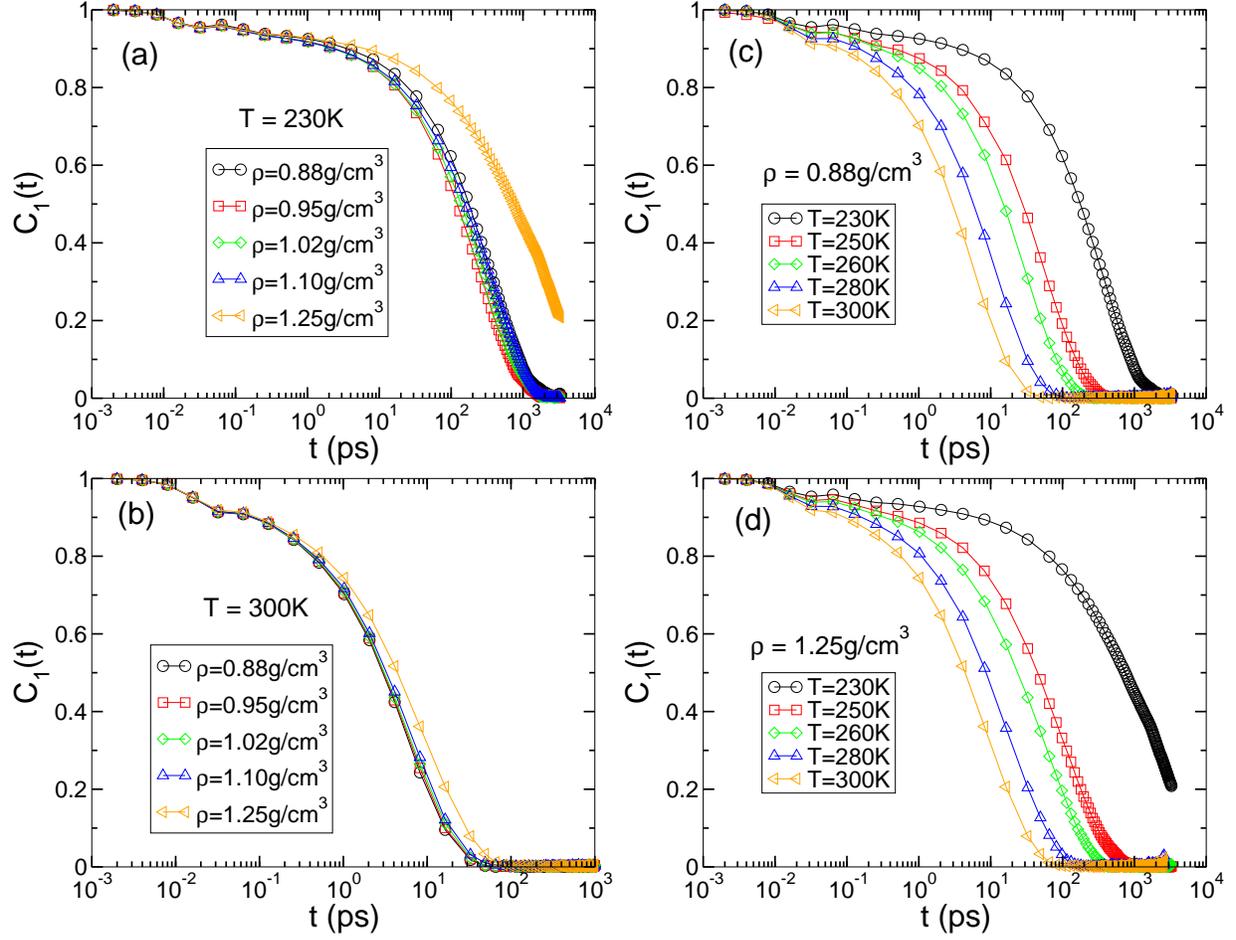

\includegraphics[width=8cm]{fig11a.eps}
\includegraphics[width=8cm]{fig11c.eps}
\includegraphics[width=8cm]{fig11b.eps}
\includegraphics[width=8cm]{fig11d.eps}
\caption{Dipole-dipole autocorrelation function $C_1(t)$ for different
   densities at constant temperature (a) $T=230$~K and (b)
   $T=300$~K. Also shown is $C_1(t)$ for different temperatures at
   constant density (c) $\rho = 0.88$~g/cm$^3$ and (d) $\rho =
   1.25$~g/cm$^3$.}
\label{ddcorr1}
\end{figure}

We next study the effect of confinement on the rotational motion of
water molecules. The rotational motion was analyzed by calculating the
rotational autocorrelation time for all the state points and are compared
with the rotational autocorrelation time for bulk water for few state
points. The rotational autocorrelation function $C_1(t)$ is defined as
\begin{equation}
C_1(t)\equiv\frac{1}{N}\left\langle\sum_{i=1}^{N}{\vec{e}_i(t).
\vec{e}_i(0)}\right\rangle,
\end{equation}
where $\vec{e}_i(t)$ is the unit dipole vector of molecule $i$ at time
$t$. For large times, $C_1(t)$ can be fit with a stretched exponential
function
\begin{equation}
C_1(t) = Ae^{(-t/\tau)^\beta}
\end{equation}
where $A$, $\beta \leq 1$ are constants and $\tau=\tau(\rho,T)$ is the
orientational autocorrelation time, which depends on both density and
temperature. In Fig.~\ref{ddcorr1}, we show $C_1(t)$ for different
temperatures and densities. Figure~12(a) shows the inverse of the
orientational autocorrelation time which is proportional to the rotational
diffusion. For a comparison with the bulk water rotational diffusion,
$\tau^{-1}$ is also shown in Fig.~12(b). Both the translational
diffusion constant [Fig.~\ref{msd2}(a)] and the inverse of the
orientational autocorrelation time [Fig.~12(a)] show similar
behavior. Similar results have been found for bulk water with the SPC/E
potential \cite{netz}. For low temperature, the maxima occur at the same
density for $D$ and $\tau^{-1}$. However at high temperatures
($T\geq260K$), where the $D$ is a monotonically decreasing function of
density, $\tau^{-1}$ has a maximum and a minimum similar to its bulk
counterpart [Fig.~12(b)].
 
\section{Crystallization of TIP5P Confined Water}
\label{sec:xtal}

Bulk TIP5P water crystallizes within the simulation time at higher densities
and, for a given density, the crystallization time has a minimum at $T =
240$~K \cite{masako}. We next investigate whether crystallization occurs in
confinement, and whether the structure differs due to the surface effects. It
has been found experimentally that water confined in hydrophobic carbon
nanopores does not crystallize, even at very low temperatures
\cite{bellisent}. However, the crystallization of confined water is seen in
some simulations \cite{zangi1,koga1}. We find that our system crystallizes to
what appears to be a trilayered ice structure at high density and that the
resultant ice has a density $\rho = 1.32$~g/cm$^3$. A similar crystallization
appears in simulations when an electric field is applied in a lateral
direction \cite{zangi2}. The surface in this kind of simulation exhibits the
embedded crystal structure of silica, where the oxygen and silicon atoms are
arranged in out-of-registry order. This suggests that the crystalline form we
find in confinement does not depend on the morphology of the surface. We show
the structure of the ice and the static lateral structure factor in
Fig.~\ref{crystal-tri}.

To investigate whether confined water crystallizes in the same
crystalline form when the separation between the plates is different, we
repeat our simulation for a plate separation of $0.7$~nm, and again the
system crystallizes at $T=260$~K. The system crystallizes into a
monolayer ice, also seen in simulations in Ref.~\cite{zangi1}. The ice
structure and its lateral static structure factor are shown in
Fig.~\ref{crystal-mono}. The density of monolayer ice is $\rho =
0.93$~g/cm$^3$ . We list the temperature, pressure, and potential energy
for these crystals in Table I.

\section{Conclusions}

We have systematically investigated the effect of confinement on TIP5P
water between two parallel smooth hydrophobic plates, separated by
$1.10$~nm. We found that the overall phase-diagram is shifted to lower
temperature and lower density compared to bulk TIP5P water by $\Delta T
\approx -40$K. The shift to lower temperature compared to the bulk water
can be understood qualitatively. Since the confinement walls do not form
any hydrogen bonds with water molecules the average number of hydrogen
bonds per molecule in confined system is smaller than in bulk
water. This is analogous to having bulk water at high temperatures.

We do not see a LL phase transition for the state points we have been
able to simulate, but we do see a pronounced inflection in the $P-\rho$
isotherms (Fig.~\ref{PhaseD}), which is consistent with a LL phase
transition at lower $T$. Since the phase diagram is shifted $\approx
40$~K lower in temperature, our results are consistent with the
possibility that there indeed is a LL transition at a temperature too
low to simulate. This shift of thermodynamics qualitatively agrees with
the theoretical predictions of Ref.~\cite{truskett}. The structure of
confined water is similar to the structure bulk water at a lower
density, and shows a similar evolution of the structure with changes in
density and temperature. At a given temperature, as the density
increases, water changes from a bilayer liquid (at low density) to a
trilayer liquid (at high density). We find that the confinement affects
the translational diffusion as well as the rotational motion of water
molecules. The rotational diffusion anomaly precedes the translational
diffusion anomaly, just as occurs for bulk water.

We were able to crystallize water for a few state points. It
crystallizes spontaneously to a trilayer ice at T=260K. Monolayer ice
was formed when the separation between the plates was decreased to
$0.7$~nm. The crystalline structures are different from the polymorphs
of bulk water, and should be relevant for confined water.

\section{Acknowledgments}

We thank C.A. Angell, E. La Nave, and F. Sciortino for discussions, NSF
grants CHE-0096892, CHE-0404699, and DMR-0427239 for support. We also
thank the Boston University Computational Center, Wesleyan University, and
Yeshiva University for supporting computational facilities.

\newpage

\begin{table}[ht]
\caption{Thermodynamic properties of the crystals formed at $T=260K$}
\begin{tabular}{c c c c}
\hline \hline

Plate Separation (nm) &   Pressure (MPa)  &   Potential Energy (kJ/mol) \\ 

0.7 & --90.30  &  --39.86  \\
1.1 &  652.89  &  --44.08  \\

\hline
\end{tabular}
\end{table}

\newpage

\newpage

\newpage

\newpage

\newpage

\newpage

\newpage

\newpage

\newpage

\newpage

\newpage

\begin{figure}[htb]
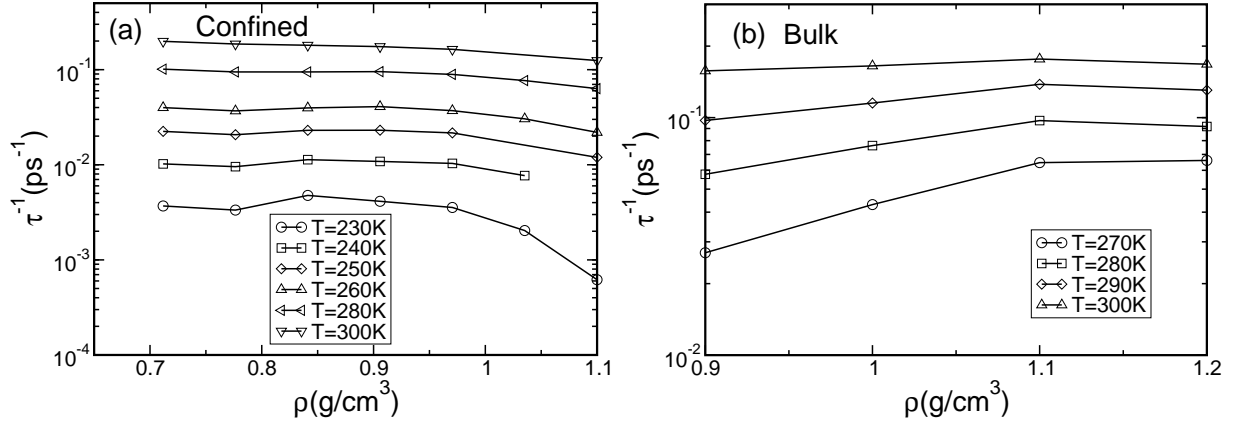

\includegraphics[width=8cm]{fig12a.eps}
\includegraphics[width=8cm]{fig12b.eps}
\caption{Inverse orientational autocorrelation time as a function of
 density for (a) confined water and (b) bulk water. Inverse
 autocorrelation time for confined water follows a similar behavior as
 bulk water, but temperature shifted by $\approx 40$~K to lower $T$.}
\label{msd3}
\end{figure}

\newpage
\eject

\begin{figure}[htb]
\includegraphics[width=12cm]{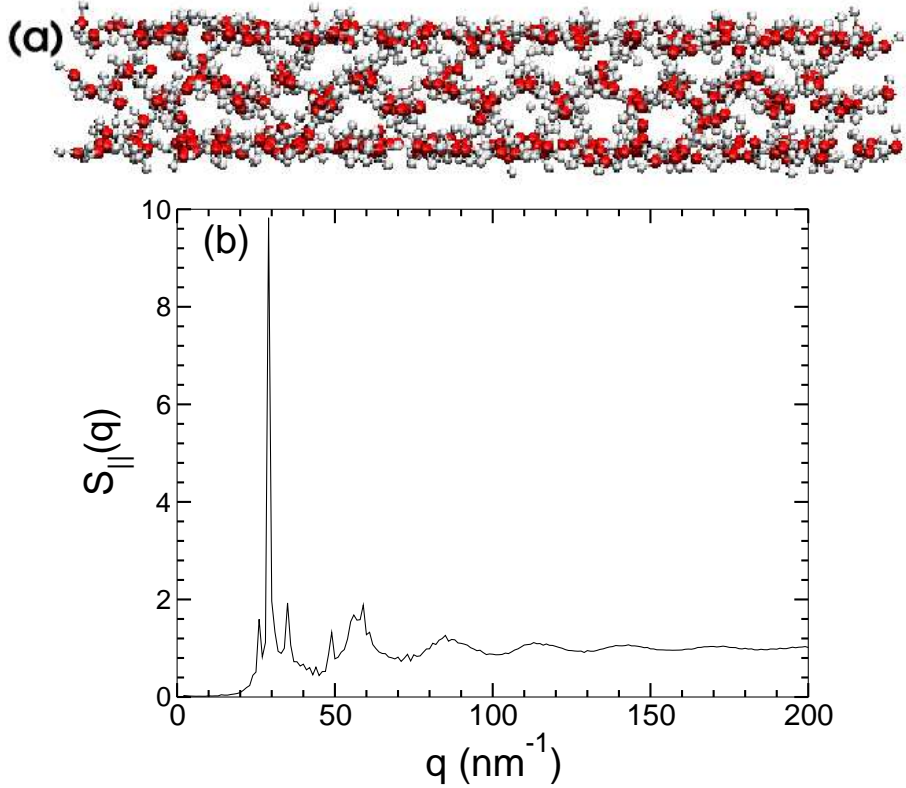} 
\includegraphics[width=10cm]{fig13b.eps}
\caption{Trilayer ice with $L_z = 1.1$~nm and $\rho = 1.32$~g/cm$^3$
  formed when the system is cooled from $T = 320$~K to $T=260$~K. (a) A
  lateral snapshot. (b) Lateral structure factor.}
\label{crystal-tri}
\end{figure}

\newpage

\begin{figure}[htb]
\includegraphics[width=12cm]{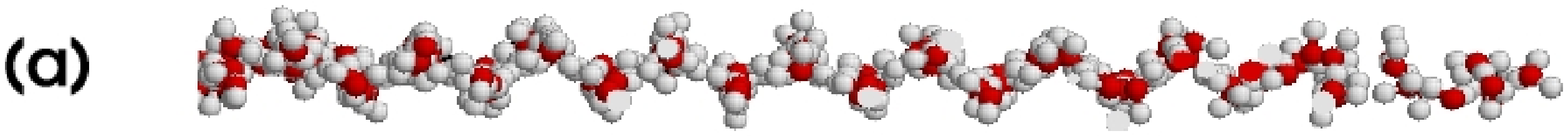}
\smallskip
\includegraphics{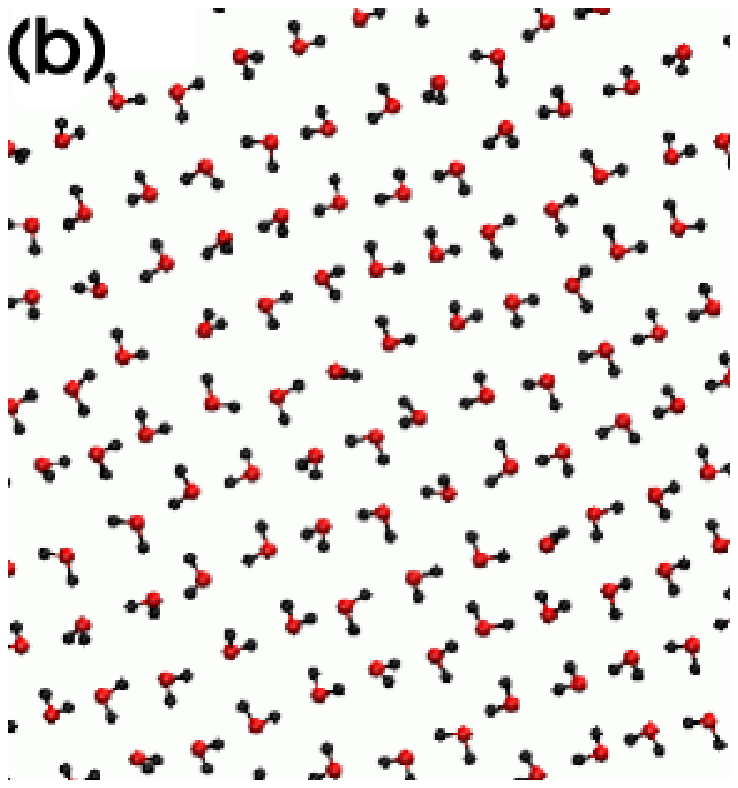}
\includegraphics[width=12cm]{fig14c.eps} 
\caption{Monolayer ice with $L_z = 0.7$~nm and $\rho = 0.93$~g/cm$^3$
  spontaneously formed at $T=260$~K. (a) A lateral snapshot and (b) top
  view. (c) Lateral structure factor. }
\label{crystal-mono}
\end{figure}


\begin{thebibliography}{99}

\bibitem{pablo-rev} C. A. Angell, Ann. Rev. Chem. {\bf 34}, 593 (1983).

\bibitem{Debenedetti03a} P. G. Debenedetti, J. Phys. Cond. Mat. {\bf
15}, R1669 (2003). 

\bibitem{Debenedetti03b} P. G. Debenedetti and H. E. Stanley, Physics Today
{\bf 56} (6), 40 (2003).

\bibitem{braskin} V. Brazhkin. S. V. Buldyrev, V. N. Ryzhov, and
H. E. Stanley, eds., {\it New Kinds of Phase Transitions:
Transformations in Disordered Substances\/} (Kluwer, Dordrecht, 2002).

\bibitem{zangi-rev}
R. Zangi, J. Phys. Cond. Mat. {\bf 16}, S5371 (2004).

\bibitem{gelb} I. D. Gelb, K. E. Gubins, R. Radhakrishnan, and
M. S. Bartkoviak, Rep. Prog. Phys. {\bf 61}, 1573 (1999).

\bibitem{chandler1} K. Lum, D. Chandler, and J. D. Weeks,
J. Phys. Chem. B {\bf 103}, 4590 (1999).

\bibitem{protein} Simulations show that the folding transition of a
protein occurs at the same time as the formation of hydrophobic protein cores in water. See R. Zhou, B. J. Berne, and R. Germain,
Proc. Nat. Acad. Sci. {\bf 98}, 14931 (2001); M. Tarek and D. J. Tobias,
Phys. Rev. Lett. {\bf 89}, 275501 (2002).

\bibitem{kb1} P. Scheidler, W. Kob, and K. Binder, Europhys. Lett. {\bf
59}, 701 (2002).

\bibitem{bellisent} M.-C. Bellissent-Funel, R. Sridi-Dorbez, and
L. Bosio, J. Chem. Phys. {\bf104}, 10023 (1996).

\bibitem{Chen95} S.-H. Chen, P. Gallo, and M.-C. Bellissent-Funel,
Can. J. Phys. {\bf 73}, 703 (1995).

\bibitem{Chen94} S.-H. Chen and M.-C. Bellissent-Funel, in {\it Hydrogen
  Bond Networks}, edited by M.-C. Bellissent-Funel and J. C. Dore, NATO
  ASI Ser.~C: Math. Phys. Sci., Vol. 435 (Kluwer Academic, Dordrecht,
  1994), p.~337.

\bibitem{gallo1} P. Gallo and M. Rovere, J. Phys. Condensed Matter {\bf
15}, 1521 (2002).

\bibitem{Spohr99} E. Spohr, C. Hartnig, P. Gallo, and M. Rovere,
  J. Mol. Liq. {\bf 80}, 165 (1999).

\bibitem{Gallo00} P. Gallo, M. A. Ricci, M. Rovere, C. Hartnig, and
  E. Spohr, Europhys. Lett. {\bf 49}, 183 (2000).

\bibitem{Hartnig00} C. Hartnig, W. Witschel, E. Spohr, P. Gallo,
  M. A. Ricci, and M. Rovere, J. Mol. Liq. (2000, in press).

\bibitem{Gallo99} P. Gallo, M. Rovere, M. A. Ricci, C. Hartnig, and
  E. Spohr, Philos. Mag. B {\bf 79}, 1923 (1999).

\bibitem{Gallo00b} P. Gallo, Phys. Chem. Chem. Phys. {\bf 2}, 1607
  (2000). 

\bibitem{pses92} P. H. Poole, F. Sciortino, U. Essmann, and
H. E. Stanley, Nature {\bf 360}, 324 (1992).

\bibitem{ms98} O. Mishima and H. E. Stanley, Nature {\bf 396}, 329
(1998).

\bibitem{poole1} P. H. Poole, F. Sciortino, T. Grande, H. E. Stanley,
and C. A. Angell, Phys. Rev. Lett. {\bf 73}, 1632 (1994).

\bibitem{speh} F. Sciortino, P. H. Poole, U. Essmann, and H. E. Stanley,
Phys. Rev. E {\bf 55}, 727 (1997).

\bibitem{slhp} H. E. Stanley, L. Cruz, S. T. Harrington, P. H. Poole,
S. Sastry, F. Sciortino, F. W. Starr, and R. Zhang, Physica A {\bf 236},
19 (1997).

\bibitem{Sciortino03}
F. Sciortino, E. La Nave, and P. Tartaglia, Phys. Rev. Lett. {\bf 91},
155701 (2003).

\bibitem{masako} M. Yamada, S. Mossa, H. E. Stanley, and F. Sciortino,
Phys. Rev. Lett. {\bf 88}, 195701 (2002).

\bibitem{hpss} S. Harrington, P. H. Poole, F. Sciortino, and
H. E. Stanley, J. Chem. Phys. {\bf 107}, 7443 (1997).

%\bibitem{brovchenko} I. Brovchenko, A. Geiger, and A. Oleinikova,
%J. Phys. Condensed Matter {\bf 16}, 5345 (2004).

%\bibitem{Stillinger97} F. H. Stillinger and D. K. Stillinger, Physica A
% {\bf 244}, 358 (1997).

\bibitem{Mishima98} O. Mishima and H. E. Stanley, Nature {\bf 392}, 164
(1998).

\bibitem{chenpaper} A. Faraone, L. Liu, C.-Y. Mou, C.-W. Yen and S.-H. Chen, 
J. Chem. Phys. {\bf 121}, 10843 (2004).

\bibitem{ChenPRVT} S.-H. Chen et al., private communication.

\bibitem{richert} S. Engemann, H. Reichert, H. Dosch, J. Bilgram,
V. Honkimaki, and A. Snigirev, Phys. Rev. Lett. {\bf 92}, 205701 (2004).

\bibitem{Ja98} E. A. Jagla, Phys. Rev. E 58, 1478 (1998);
E. A. Jagla, J. Chem. Phys. 110, 451 (1999);
E. A. Jagla, J. Chem. Phys. 111, 8980 (1999).

\bibitem{Bul02} S. V. Buldyrev, G. Franzese, N. Giovambattista,
G. Malescio, M. R. Sadr-Lahijany, A. Scala, A. Skibinsky, H. E. Stanley,
Physica A 304, 23 (2002).

\bibitem{Fr01} G. Franzese, G. Malescio, A. Skibinsky, S. V. Buldyrev
and H. E. Stanley, Nature 409, 692 (2001).

\bibitem{fran02} G. Franzese, G. Malescio, A. Skibinsky, S. V. Buldyrev
and H. E. Stanley, Phys. Rev. E 66, 051206 (2002).

\bibitem{Ku04}P. Kumar, S. V. Buldyrev, F. Scioritno, E. Zaccarelli,
H. E. Stanley, Phys. Rev. E (in press) and cond-mat/0411274 (2004).

%\bibitem{sfa2002} S. V. Buldyrev, G. Franzese, N. Giovambattista,
%G. Malescio, M. R. Sadr-Lahijany, A. Scala, A. Skibinsky, and
%H. E. Stanley, Physica A {\bf 304}, 23 (2002).
 
%\bibitem{my-preprint} M. Yamada, et al., unpublished.

%\bibitem{se2003} S. V. Buldyrev and H. E. Stanley, Physica A {\bf 330},
 %124 (2003).

\bibitem{jorgensen1} M. W. Mahoney and W. L. Jorgensen,
J. Chem. Phys. {\bf 112}, 8190 (2000).

\bibitem{jorgensen2} M. W. Mahoney and W. L. Jorgensen,
J. Chem. Phys. {\bf 114}, 363 (2001).

\bibitem{paschek1}
D. Paschek, arXiv:cond-mat/0411724.

\bibitem{meyer} M. Meyer and H. E. Stanley, J. Phys. Chem. B {\bf 103},
9728 (1999).

\bibitem{tip4p} W. L. Jorgensen, J. Chandrasekhar, J. Madura, R. W. Impey
and M. Klein, J. Chem. Phys. {\bf 79}, 926 (1983).

\bibitem{tip4p-tanaka1} H. Tanaka, Nature, {\bf 380}, 6572 (1996).

\bibitem{tip4p-tanaka2} H. Tanaka, J. Chem. Phys. {\bf 105}, 5099 (1996).

\bibitem{koga2} K. Koga, H. Tanaka, and X. C. Zeng, Nature, {\bf 408},
564(2000).

\bibitem{koga3} K. Koga and H. Tanaka, J. Chem. Phys. {\bf 122}, 104711 (2005).

\bibitem{truskett} T. M. Truskett and P. G. Debenedetti,
J. Chem. Phys. {\bf 114}, 2401 (2001).

%\bibitem{csss} M. Canpolat, F. W. Starr, M. R. Sadr-Lahijany, A. Scala,
%O. Mishima, S. Havlin, and H. E. Stanley, Chem. Phys. Lett. {\bf 294}, 9
%$(1998).

%\bibitem{as2004} A. Skinbinsky, S. V. Buldyrev, G. Franzese,
%G. Malescio, and H. E. Stanley, Phys. Rev. E {\bf 69}, 061206 (2004).

%\bibitem{pgdbook} P. G. Debenedetti, {\it Metastable Liquids: Concepts
% and Principles}, (Princeton University Press, Princeton, 1998).

%\bibitem{glosli1998} J. N. Glosli and F. H. Ree, Phys. Rev. Lett. {\bf
%82}, 4659 (1999).

\bibitem{koga1} K. Koga, X. C. Zeng, and H. Tanaka,
Phys. Rev. Lett. {\bf 79}, 5262 (1997).

\bibitem{zangi1} R. Zangi and A. E. Mark, Phys. Rev. Lett. {\bf 91},
0255502 (2003).

\bibitem{zangi2} R. Zangi and A. E. Mark, J. Chem. Phys. {\bf 120}, 7123
(2004).

\bibitem{tanaka2} J. Slovak, K. Koga, H. Tanaka, and X.C. Zeng,
Phys. Rev. E {\bf 60}5833 (1999).

\bibitem{st2} F. H. Stillinger and A. Rahman, J. Chem. Phys. {\bf 60},
1545 (1974).

\bibitem{sorenson} J. M. Sorenson, G. Hura, R. M. Glaeser, and
T. Head-Gordon, J. Chem. Phys. {\bf 113}, 9149 (2000).

\bibitem{Ohmine} Matsumoto M, Saito S, and Ohmine I., Nature, {\bf416} 6879(2002).

\bibitem{baez} L. A. Baez and P. Clancy, J. Chem. Phys. {\bf 103}, 9744
(1995).

\bibitem{lee2} C. Y. Lee, J. A. McCammon, and P. J. Rossky,
J. Chem. Phys. {\bf 80}, 4448 (1984).

\bibitem{hansen} J. P. Hansen and I. R. McDonald, {\it Theory of Simple
Liquids}, (Academic Press, London, 1996).

\bibitem{lee1} S. H. Lee and P. J. Rossky, J. Chem. Phys. {\bf 100},
3334 (1994).
 
\bibitem{footnoteRealLz} We note the the accessible volume to the water
molecules is not given by the total geometrical volume between the walls,
$V=L_xL_yL_z$. Instead, the accessible volume is given by
$V'=L_xL_yL'_z$ where $L'_z<L_z$ due to the repulsive wall-water
interactions for short distances. The calculation of $L'_z$ is explained
in Sec.~\ref{sec:rhoz}.

\bibitem{berend} H. J. C. Berendsen, J. P. M. Postma, W. F. van
Gunsteren, A. DiNola and J. R. Haak, J. Chem. Phys. {\bf 81}, 3684
(1984).

\bibitem{virial} M. P. Allen and D. J. Tildesley, {\it Computer
Simulation of Liquids\/} (Oxford University Press, 1997).

\bibitem{finneyPRL} J. L. Finney, A. Hallbrucker, I. Kohl, A.K. Soper,
and D.T. Bowron, Phys. Rev. Lett. {\bf 88}, 225503 (2002).

\bibitem{soperRicci} A.K. Soper and M.A. Ricci, Phys. Rev. Lett. {\bf
84}, 2881 (2000).

\bibitem{sslsss} F. W. Starr, S. Sastry, E. La Nave, A. Scala,
  H. E. Stanley and F.  Sciortino, Phys. Rev. E {\bf 63}, 041201 (2001).

\bibitem{francis} F. W. Starr, S. T. Harrington, F. Sciortino, and
  H. E. Stanley, Phys. Rev. Lett. {\bf 82}, 3629 (1999); F. W. Starr,
  F. Sciortino, and H. E. Stanley, Phys. Rev. E {\bf 60}, 1084 (1999).

\bibitem{benedek}
G. Benedek and F. Villars, {\it Physics with illustrative examples from
medicine and biology} (Addison Wesley, Reading, 1975)

\bibitem{sns} F. W. Starr, J. K. Nielsen, and H. E. Stanley,
Phys. Rev. Lett. {\bf 82}, 2294 (1999); Phys. Rev. E {\bf 62} 579
(2000).

\bibitem{netz} P. A. Netz, F. Starr, M. C. Barbosa, H. E. Stanley,
J. Mol. Liq. {\bf 101}, 159 (2002); P. A. Netz, F. W. Starr,
H. E. Stanley, and M. C. Barbosa, J. Chem. Phys. {\bf 115}, 344--348
(2001).

\end{thebibliography}
\end{document}